# DEWPython: A Python Implementation of the Deep Earth Water Model and Application to Ocean Worlds

ANDREW CHAN[1,2], MOHIT MELWANI DASWANI[2*], STEVEN VANCE[2],

[1]Div. of Geological and Planetary Sciences, California Institute of Technology, Pasadena, CA, USA 91125

[2] Jet Propulsion Laboratory, California Institute of Technology, M/S 183-301 4800 Oak Grove Dr., Pasadena, CA 91109, USA

*Corresponding author (mohit.melwani.daswani@jpl.nasa.gov)

**Link to code:**

https://github.com/chandr3w/DEWPython

**Authorship Statement**

Chan coded DEWPython and wrote the manuscript. Melwani Daswani aided in writing the manuscript, resolving coding issues, provided scientific motivation, and revised the manuscript. Vance served as the advisor to the project and revised the manuscript.

**Abstract:**

There are two main methods of calculating the thermodynamic properties of water and solutes: mass action (including the Helgeson-Kirkham-Flowers (HKF) equations of state and model) and Gibbs free energy minimization (e.g. Leal et al., 2016). However, in certain regions of pressure and temperature the HKF model inaccurately predicts the speciation and concentration of solutes (e.g. Miron et al., 2019). The Deep Earth Water (DEW) model uses a series of HKF-type equations to calculate the properties of water and solute concentrations at high temperatures (373 – 1473 K) and pressures (0.1 – 6 GPa) (e.g. Huang and Sverjensky, 2019; Pan et al., 2013; Sverjensky et al., 2014). The DEW model is synthesized in an Excel spreadsheet and calculates Gibbs energies of formation, equilibrium constants, and standard volume changes for reactions. Here we present an object-oriented Python implementation of the DEW model, called DEWPython. Our model expands on DEW by increasing model efficiency, streamlining the input process, and incorporating SUPCRT in-line. Additionally, our model builds in minerals and aqueous species from the thermodynamic database slop16.dat (Boyer, 2019) which would normally be calculated separately. We also present a set of reactions relevant to icy ocean world interiors calculated with the DEWPython. The favorability of these reactions indicates likely formation of certain organic species under extreme pressures relevant to ocean worlds.

## 1. Introduction

### *1.1. The need for improved geochemical frameworks to ocean worlds*

Internal liquid water beneath the ice shells of ocean worlds, including Europa, Enceladus, and Titan could be suitable for life (e.g. Mann, 2017). The differing input materials, thermal histories, and geophysical properties of these moons (e.g. radius, density) likely results in differing ocean compositions (e.g. Vance et al., 2018, Neri et al. 2020). Titan and Europa represent endmember ocean worlds with different ocean compositions and physical properties.

Titan is an icy moon with an organic-rich (containing methane and other hydrocarbons) atmosphere and is the only moon in the solar system with any significant atmosphere. The surface temperature is 94 – 97 K and additionally has a methane-centric cycle, similar to that of water on earth (Jennings et al., 2016). Although there are several hypotheses, the exact origin of the atmosphere on Titan is uncertain (Owen, 2000). One hypothesis is that the atmosphere on Titan originated from its ocean (e.g. Miller et al., 2019). If supported, this hypothesis would have direct implications for the evolution of Titan, the current oceanic composition, and the potential habitability of its ocean. A methane-rich ocean, for example, would result in a lower freezing temperature compared to a pure water ocean.

Europa's ocean similarly has an uncertain history of formation (e.g., Pappalardo, 1999). Based on spectroscopic observations, a few main hypotheses exist about the composition of the modern ocean, where it could contain hydrated sulfide or sulfate salts, chloride salts, or some combination thereof (Dalton et al., 2013; Fischer et al., 2016; Ligier et al., 2016; Trumbo et al., 2019). The salt composition and concentration could potentially constrain the temperature of Europa's ocean as well as the potential habitability (Johnson et al. 2019, Buffo et al. 2020, Vance et al. 2021).

Reliable geochemical modelling is necessary to determine which of these hypothesized compositions is the most robust. The semi-empirical Helgeson-Kirkham-Flowers (HKF) equation of state is in broad current use for computing the thermodynamic properties of aqueous complexes and ions at differing temperatures and pressures. However, HKF provides unphysical results in certain regions of pressure-temperature space where it is inaccurate at predicting the properties of water (e.g., density, Gibbs energy) due to inadequate experimental data on the dielectric constant of water at high temperatures and pressures (Miron et al., 2019). The improvement of this model and the equations of state for organic molecules at various pressures and temperatures can help inform about the rates at which they form, their mechanisms of formation, and the locations at which they form.

With planned NASA missions Europa Clipper and Dragonfly to Europa and Titan respectively, the accuracy and usability of thermodynamic models need to be improved; better models enable the prediction of chemical processes occurring in these worlds and allow preparation for the analysis of spacecraft measurements, such as ocean compositions inferred from molecular species possibly present in Europa's exosphere or Titan's atmosphere.

### *1.2. Software Background*

Several thermodynamic frameworks calculate thermodynamic properties of aqueous systems at extreme ranges of temperature and pressure relevant to icy ocean worlds, with different relative merits. We review them here to provide context for the development of DEWPython.

SUPCRTBL (Zimmer et al., 2016) is a refinement of SUPCRT92 (Johnson et al., 1992) which calculates the Gibbs energy of formation, volume of formation, and changes in entropy, enthalpy, and specific heat for minerals, aqueous species, and reactions. Also pertinent to this work, SUPCRT96 is an intermediary between SUPCRT92 and SUPCRTBL that incorporates updated equations, however, still produces values in the units of SUPCRT92. These properties are calculated either under conditions where the pressure is equivalent the liquid-vapor saturation conditions (for our use, 273–623 K, 1–16.2 MPa– i.e. *psat*), at isobars and isotherms ranging from 273–1273 K and 0–500 MPa, or at any point within a P-T grid.

SeaFreeze (Journaux et al., 2020) is a local basis function approach to calculating the properties of icy polymorphs Ih, III, V, and VI, at the complete range of conditions that could be found in oceans and hydrospheres in our solar system (200 – 400 K, 0 – 2300 MPa). The temperature range of SeaFreeze is also extendable past this upper end for the properties of liquid water only. The density is calculated in SeaFreeze as a derivative of Gibbs energy with pressure that is accurate in the high pressure—temperature (PT) range.

The Deep Earth Water model (DEW; Huang and Sverjensky, 2019; Pan et al., 2013; Sverjensky, 2019; Sverjensky et al., 2014) is synthesized in an Excel spreadsheet that similarly calculates Gibbs energy of formation, equilibrium constants, and standard volume change for reactions. DEW has an extended range of validity and can function under several other equations of state (general validity ~ 373 – 1473 K, 100 – 6000 MPa). There are several online implementations of DEW, such as in ENKI ThermoEngine (Ghiorso and Wolf, 2019) and CHNOSZ (Dick, 2019), however, there is no specific one-to-one replication of the original spreadsheet in Python format (ThermoEngine can run DEW reactions, but does not have the flexibility of the original model).

The main goals of this work are to make the outputs from DEW transferrable between programs and to speed up calculations of mineral species (normally the properties of each mineral species would have to be calculated externally and then input into DEW). The Python implementation (Supplementary Text S1) allows for a direct comparison of different thermodynamic calculations and a Python-centric evaluation of reactions at extreme temperatures and pressures. DEWPython additionally includes minerals and aqueous species from the thermodynamic database slop16.dat (Boyer, 2019) typically known as a SUPCRT dataset.

## 2. Methods

We reconstructed DEW as the open-source Python package DEWPython with the goal of combining a wide P-T range and the calculation of joint mineral and organic thermodynamic properties and aqueous speciation. Table 1 compares the general features of the thermodynamic models synthesized into DEWPython.

Table 1: Features of thermodynamic models compared to DEWPython.

| | DEW | SUPCRT | SeaFreeze | DEWPython |
|---|---|---|---|---|
| **Robust water properties** | | | X | X |
| **Python-integrable** | | | X | X |
| **Integrated organic calculations** | X | X | | X |
| **Integrated mineral calculations** | | X | | X |
| **Pressure range 0.1 – 6 GPa** | X | | X | X |
| **Temperature range 373.15 – 1473.15 K** | X | | X | X |

### *2.1. Code conversion*

The base main code of DEWPython converts the functions from the Deep Earth Water Model's VBA storage into Python functions. This porting was accomplished with the aid of the Vb2Py Python module (http://vb2py.sourceforge.net/). These functions were then further rewritten to remove the dependency on Vb2Py. An internal inconsistency in the representation of the constant "G" was corrected in DEWPython. Converted functions were tested directly against their Excel counterparts.

To reconstruct the interface for DEW (see Supplementary Text S1), input/validation loops were utilized and combined into the "set_inputs()" and "set_outputs()" methods of the DEWModel class. The custom sheets for water density, dielectric constant, and water free energy were integrated using the comma separated value format using the "import_custom_sheets()" method. To mimic the "input" sheet of DEW, functions "set_TPRho() and set_preferences()" were coded using a similar interactive input/validation loop.

The embedded dictionaries for aqueous species and gaseous species were created utilizing the pandas package to upload the DEW spreadsheet and store every species in text files. To create the embedded mineral dictionary, every mineral in the slop16 thermodynamic data file (Boyer, 2019) was run through SUPCRT96 under *psat* conditions. The lower limit on temperature was set at 25 °C as the *psat* pressure does not change until above 100 °C. Additionally, SUPCRT96 was used instead of SUPCRTBL due to the unique ability to handle the slop16 data file. These SUPCRT output txt files were manually combined using a custom file input code and created as a nested dictionary of thermodynamic properties for each mineral species available.

The equations for each cell are handled in the "calculate_aq()", "calculate_gas()", and "calculate_H2O()". The individual calculations for DEW were handled with numpy arrays substituting the Excel sheet columns, and the equations are directly substituted from the spreadsheet, pulling the values that would be on the "Reaction Information" spreadsheet directly from the embedded dictionaries. Each of the calculate functions was written to update input and output arrays for aqueous, gaseous, and water inputs/outputs. The "calculate()" function runs each these submethods and combines the values into single arrays of "delG", "logK", and "delV"/ If mineral inputs/outputs are being used, their values will be queried from the dictionary during the "calculate()" function and subsequently combined with these arrays as well.

Plots from the Excel spreadsheet's "Results" sheet were replicated using the "make_plots()" method. This method uses sets and default dictionaries to sort the pressure and temperature lists and create tuples of pressure and temperature combinations to properly separate the plots by temperature and pressure increments.

SUPCRTBL was implemented in the function using the subprocess module of Python. The "outLoop()" function contains and prints the output of SUPCRTBL that would otherwise be in the command line. The "run_supcrt()" method utilizes "subprocess.Popen()" and a "while" loop to interactively take input and to end when the text file from SUPCRTBL is produced. Note, this method requires the user to save reactions in the ".rxn" file format.

The last two methods implemented were the "calculate_supcrt()" and the "calculate_supcrt_special()" methods. The former will calculate the outputs from a *psat* SUPCRT96 or SUPCRTBL file and store it as a nested dictionary of reactions. The latter method performs the same calculations and storage. However, this method performs the calculations for files calculated at isotherms and isobars beyond *psat* temperatures. The "make_supcrt_plots()" method creates the same plots as DEW for pre-input SUPCRTBL or supcrt96 fields. This method currently only handles the plots for the *psat* conditions.

The code for DEWPython was uploaded and published on Github with the inclusion of additional __init__, manifest, license, and setup files. This package was then uploaded to the Python Package Index (PyPI) utilizing the sdist/bdist_wheel Python functions and the twine utility. Documentation for the DEWPython package was created using the pdoc3 package (https://pdoc3.github.io/pdoc/) and was provided through Github pages.

We validated the model through direct comparisons the model through direct comparisons of reactions and species to the values output by the DEW spreadsheet. Following the validation of the model, the heretofore mentioned reactions were run in Jupyter notebooks. The same reactions were run in SUPCRT96 and integrated into a DEW() object through the "calculate_supcrt()" method. The comparison plots were created through 3D plotting in matplotlib allowing simple and direct comparisons between DEW and supcrt. An example of the validation between the DEW spreadsheet and DEWPython is shown in Supplementary Figures S1 – S2. The SeaFreeze values were calculated separately implementing SeaFreeze at the same pressures and temperatures as a pure reaction of a single water molecule.

The integrated SUPCRT96 in the DEWPython package was used to run the suite of reactions at a PT range of 673 – 1473 K (the upper range of DEW) and 100 – 500 MPa (the upper range of SUPCRT96).

### *2.2. Model validation and reactions relevant to icy ocean worlds*

We modeled a number of reactions with key significance to Europa and Titan in order to compare the differences between the output of each thermodynamic model incorporated into DEWPython, and to investigate their relevance in the oceans of Europa and Titan. Specifically, we investigated the stabilities of methanol, methane and ethane, and calcite. The Cassini mission determined that there is both ethane and methane in Titan's atmosphere (e.g. Griffith et al., 2006). The stability of these substances in the ocean could either validate or rule it out as a possible source of atmospheric methane/ethane, as opposed to photochemical or meteoritically delivered. Methanol potentially acts as an antifreeze in Titan's ocean (Dougherty et al., 2018). Calcite is a carbon sink on Earth and a key component of some living organisms; the presence and stability of carbonate minerals in icy ocean worlds could be indicative of its availability to possible organisms, as well as the redox state of carbon (Glein and Waite, 2020).

We additionally modeled the stability of acetic acid and pyruvate, which are listed in the Supplementary Material (Supplementary Figures S3 – S9). These two compounds are also relevant to prebiotic chemistry (e.g. Barge et al., 2020, 2019).

## 3. Results and Discussion

In our comparison of DEWPython to SeaFreeze, DEW, and SUPCRT96, we find that generally DEWPython is consistent with these models. Specifically DEWPython produces identical results as DEW, results within 1% of SUPCRT when the same source for aqueous/gaseous species is used, and within 3% of SeaFreeze at valid ranges. As such, DEWPython can accurately predict properties of aqueous species under conditions of elevated pressure and temperature occurring in icy ocean worlds.

### *3.1. Comparison of water properties between DEW and SeaFreeze*

The density of water calculated by SeaFreeze and that calculated by DEWPython ($\rho$ SeaFreeze – $\rho$ DEWPython) agree within a difference of ~ 2 % (Figure 1). At lower temperatures and pressures this difference is less than 1 %, however, at the temperature and pressure extremes (~6000 MPa, 1300 K) calculated in this study the differences reach slightly over 2%. DEWPython is inaccurate at low temperatures and high pressures (>3000 MPa, 373-573 K) as this region falls in the range of ice VII which is not accounted for in the model.

The difference in the Gibbs free energy of the formation of water ($\Delta G_f$ SeaFreeze – $\Delta G_f$ DEWPython; Figure 2) tends to increase between DEWPython and SeaFreeze with increasing temperature.

SeaFreeze calculates nearly identical values for water density as the Deep Earth Water model (DEW) does. However, the two differ slightly at high pressures and temperatures, as expected from the parameterization of DEW using the Zhang and Duan (2005) equation for the density of water (SeaFreeze "water2" uses the Brown 2018 equation of state which uses local basis functions). At high pressures and temperatures, the Zhang and Duan equation of state becomes less reliable and is only valid up to the 1473 K limit. Given that the Gibbs free energy derives directly from water density in DEW, calculated Gibbs free energy exhibits similar behavior. The implication of this result is that DEW may not actually need to be re-evaluated in context of differing Gibbs of water values for DEW's range of validity.

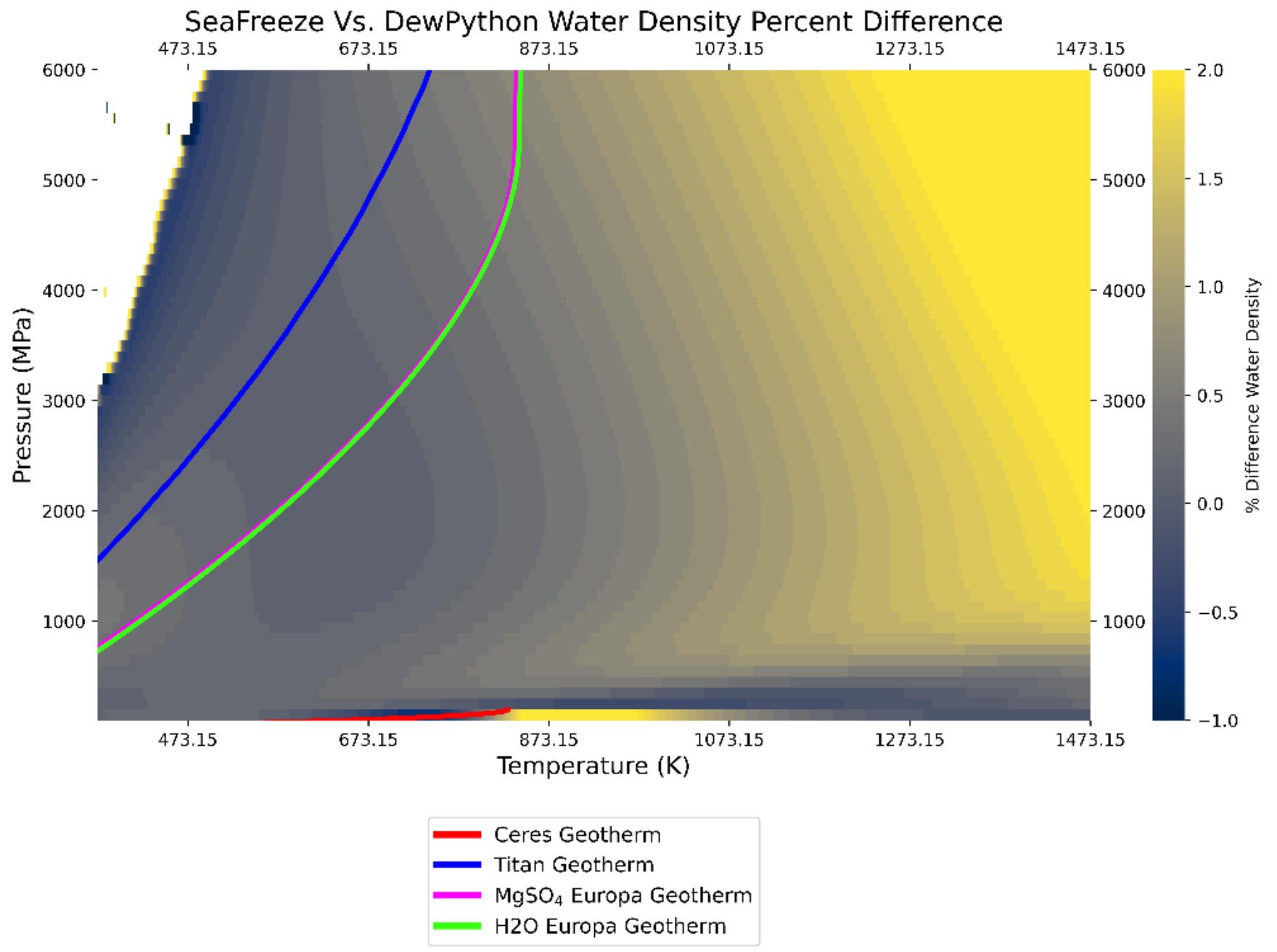

*Figure 1. Percent difference in water density between DEW and SeaFreeze from 373 – 1300 K and 100 – 6000 MPa. Ceres geotherm is from Castillo-Rogez and McCord (2010), Titan and Europa geotherms are from Vance et al. (2018). Areas with >5% difference are whited out: these fall in/near the range of ice VII.*

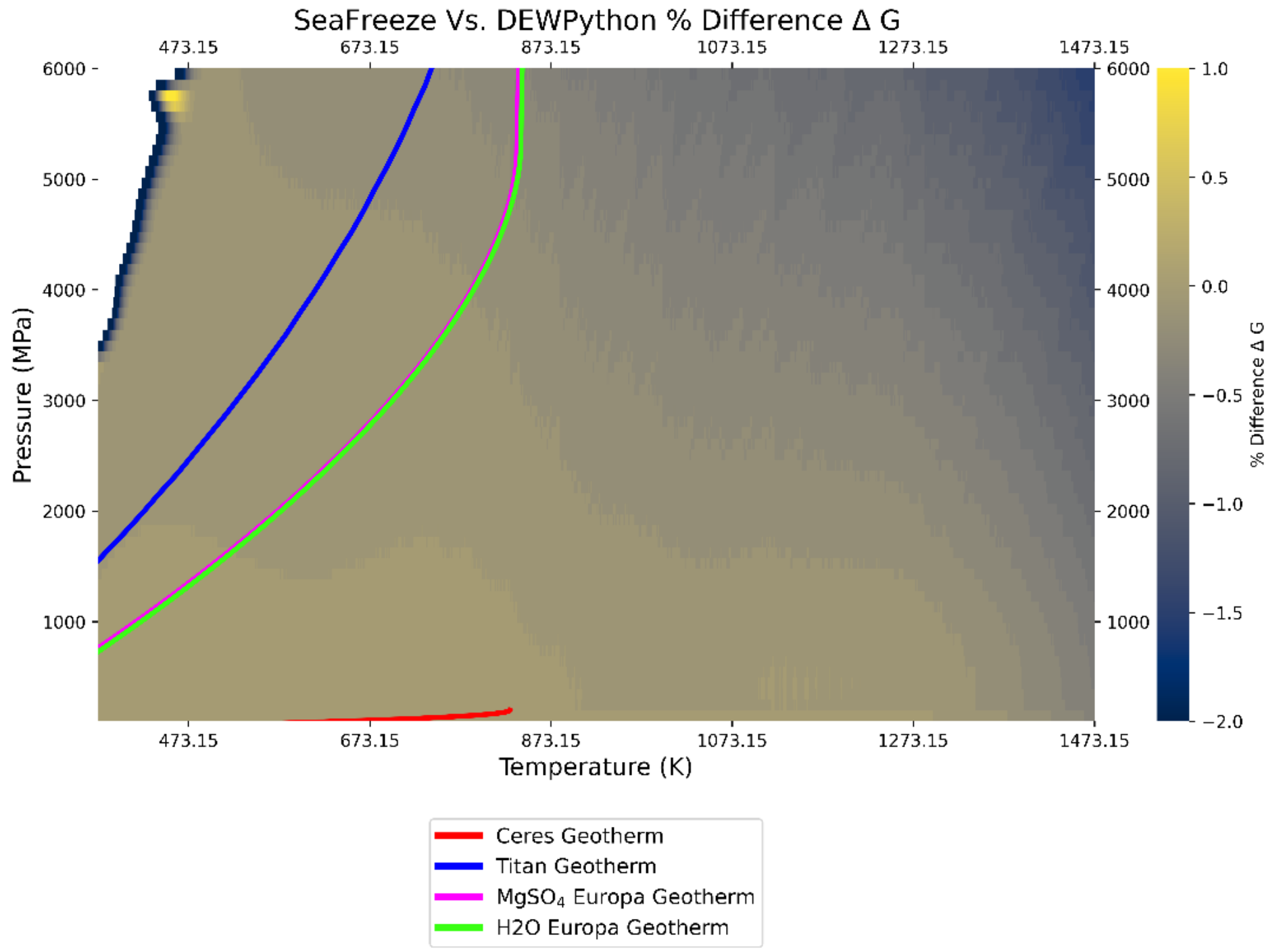

*Figure 2. Percent difference in Gibbs energy of formation between DEW and SeaFreeze from 373 – 1300 K and 100 – 6000 MPa. Ceres geotherm is from Castillo-Rogez and McCord (2010), Titan and Europa geotherms are from Vance et al. (2018). Areas with >5% difference are whited out: these fall in/near the range of ice VII.*

## 3.2. Comparison between DEW and SUPCRT reactions

### A. Methanol

The stability of methanol ($CH_3OH$) is strongly dependent on the specific reaction pathway and the aqueous reactants and products. We focus on the dissociation of methanol into the methyl and hydroxyl radicals ($CH_3OH(aq) \rightleftharpoons CH_3(aq) + OH(aq)$) in the wide pressure-temperature (PT) space enabled by SUPCRT and DEW, and show other possible methanol formation and dissociation pathways in the Supplementary Material (Supplementary Figure S10). Figure 3 shows the equilibrium constant (in standard $\log_{10}$ mol units) for the dissociation reaction in the available PT space. Methanol is relatively stable compared to its dissociation products, particularly in the high temperature, low pressure regime. Comparing the stability of the reaction across different bodies by overlaying the geotherms of Titan, Europa and Ceres obtained from the literature, methanol appears to be most stable within the interior of Ceres, becoming more stable with increasing depth. The "warm" Europa geotherm is most conducive to the dissociation of methanol, but even in that case, methanol is still more stable relative to its dissociation products. Should methanol be present in the interiors or oceans of these bodies, it appears likely that it would not dissociate unless other reactants would cause it to do so (e.g. oxidants).

In the overlapping PT regions between SUPCRT and DEW, calculated log Ks differ somewhat between the models, particularly at higher pressures and temperatures. For example, following the geotherm for Ceres, SUPCRT yields log K $<-5$ at the highest pressure within Ceres, whereas DEW yields log K $>-4$.

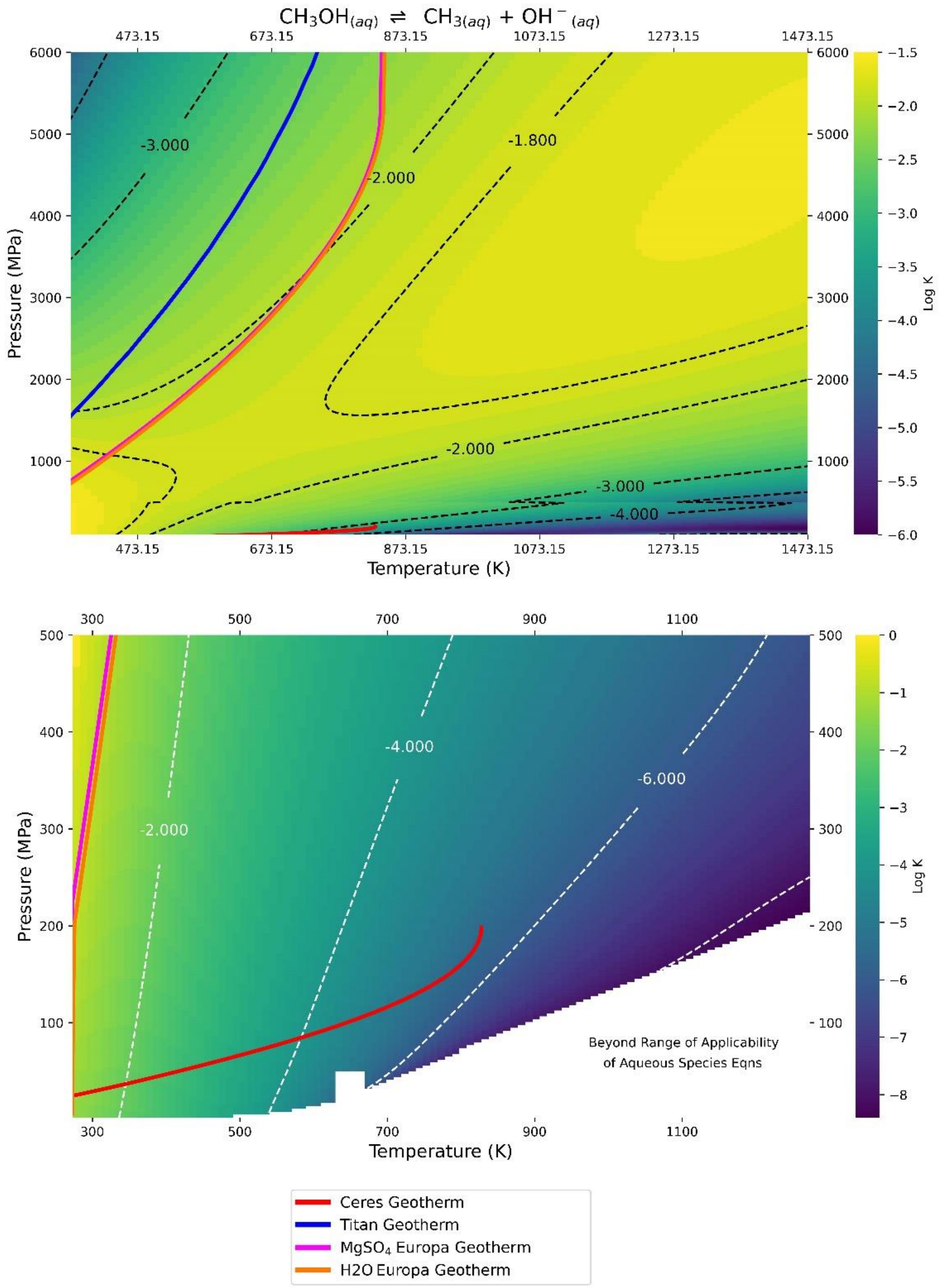

*Figure 3. Equilibrium constant values ($\log_{10}$ K) calculated for the dissociation of methanol $CH_3OH \rightleftharpoons CH_3 + OH^-$. Ceres geotherm is from Castillo-Rogez and McCord (2010), Titan and Europa geotherms are from Vance et al. (2018). a) log K for the PT range of DEW. b) log K for the PT range of SUPCRT.*

*B. Ethane and methane*

As discussed, Titan's atmosphere contains both ethane and methane. Here we explore the stability of ethane in water $C_2H_6 + H_2 \rightleftharpoons 2CH_4$; other reaction pathways and extended calculations are shown in Supplementary Figures S11 – S13. The formation of methane from ethane and hydrogen is favorable under pressure and temperature conditions relevant to Titan's and Europa's oceans (Fig. 4), suggesting that ethane formation in their oceans through this pathway is inhibited. The reaction only becomes unfavorable in the deeper interiors, i.e., in the silicate mantles of Europa and Titan. If ethane converts to methane in Titan's ocean, it would imply that the ethane discovered by Cassini in Titan's atmosphere would have originated from photochemistry in the atmosphere, surface or ice shell, or have an external source. Ethane additionally could be produced by surface lakes or directly in the atmosphere utilizing a cyclical process where evaporation or volcanic action release methane that converts to ethane. Minor organic solutes $C_2H_6$, $CH_4$ and $N_2$ evaporate to $CH_4$ in the atmosphere. This would combine with $N_2$ to form complex organics/will also form $C_2H_6$ in the atmosphere, which precipitate back into the lake through aerosol sedimentation (Raulin, 2008). Alternatively, ethane production from methane could occur within the silicate interior of Titan, although Titan's interior is unlikely to be hydrated (e.g. Monteux et al., 2018).

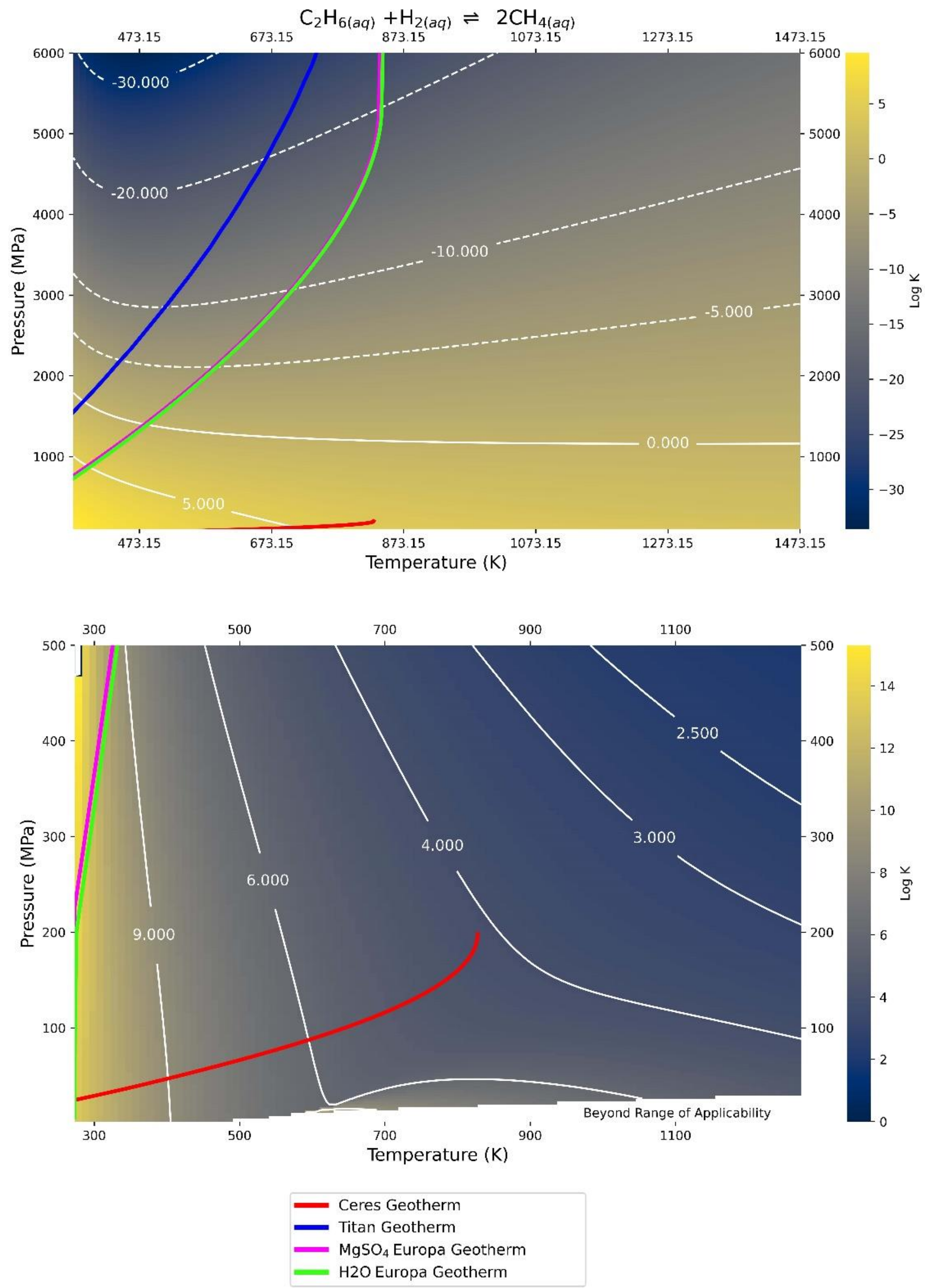

*Figure 4. Equilibrium constant values ($\log_{10}$ K) calculated for the reaction $C_2H_6 + H_2 \rightleftharpoons CH_4$. Ceres geotherm is from Castillo-Rogez and McCord (2010), Titan and Europa geotherms are from Vance et al. (2018). a) log K for the PT range of DEW. b) log K for the PT range of SUPCRT.*

### C. Aqueous calcium carbonate

Aqueous calcium carbonate formation from its constituent ions ($Ca^{+2} + CO_3^{2-} \rightleftharpoons CaCO_3$) is favorable throughout the P-T space encompassed by SUPCRT and DEW, but is highest at the low pressure-high temperature regime, and high pressure-low temperature regime (Fig. 5). The formation of aqueous calcite is most favorable in the deep interior of Ceres (log K > 8; Fig. 5).

This result indicates that the formation of calcium carbonate is potentially an effective carbon sink in icy ocean worlds: once formed, carbon stored as calcite would not be easily soluble in icy ocean worlds (similar to its behavior on Earth e.g., Broecker & Takahashi, 1977). It is also important to note that the log K values calculated by SUPCRT and DEW diverge at high temperature values (Fig. 5). This difference partially results from the different equations; however, the bulk of the difference likely arises from a difference in the source of thermodynamic data. Specifically, DEW uses aqueous calcium carbonate data from Facq et al. (2014) and the slop16 thermodynamic file uses calcium carbonate data from Sverjensky et al. (1997) (and generally contains aqueous species data from several different sources).

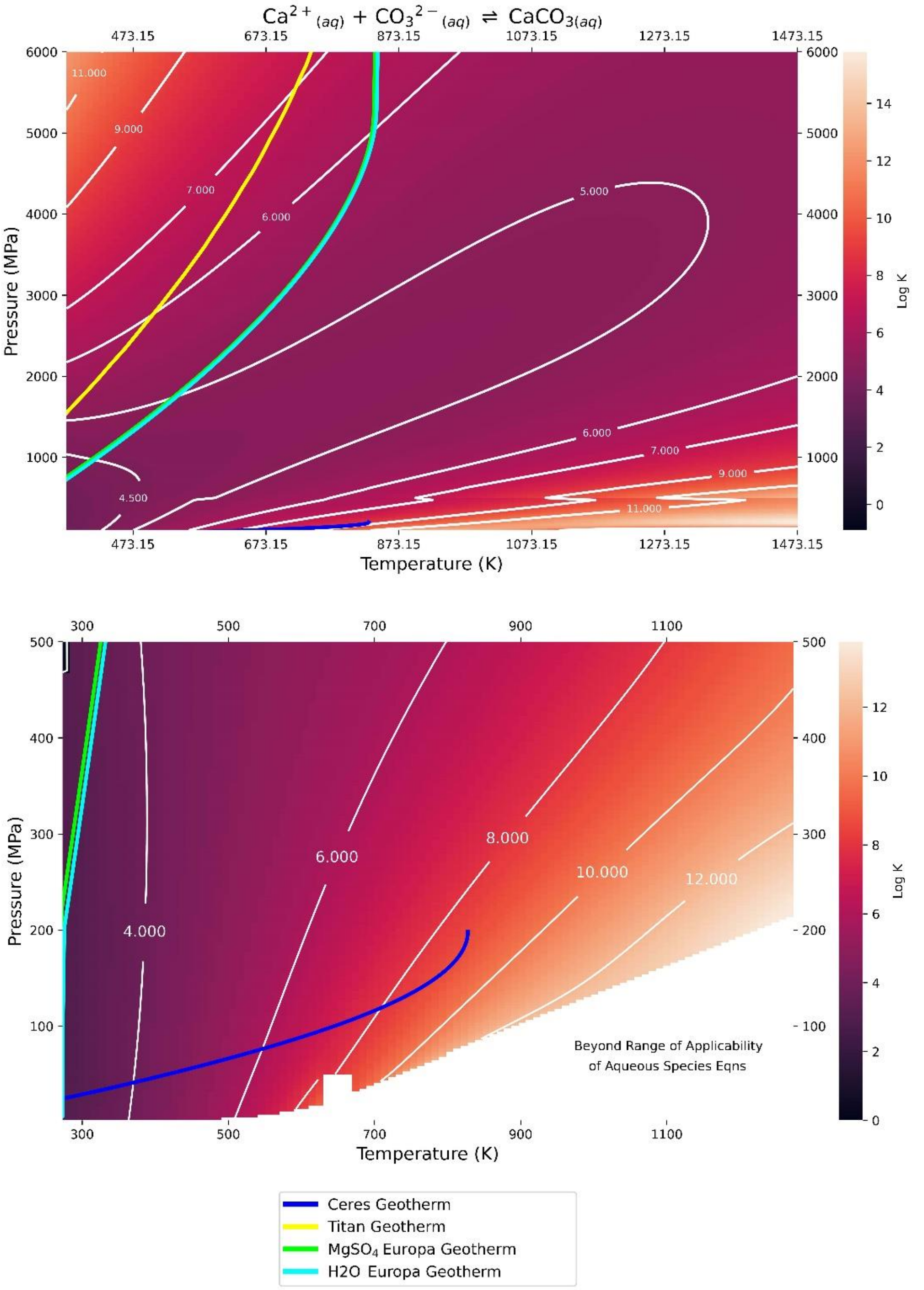

*Figure 5. Equilibrium constant values (log K) calculated for the formation of aqueous calcium carbonate, $Ca^{2+} + CO_3^{2-} \rightleftharpoons CaCO_3$. Ceres geotherm is from Castillo and McCord (2010), Titan and Europa geotherms are from Vance et al. (2018).*

## 4. Conclusion and Future Directions

We have created new functionality for the Deep Earth Water Model by implementing it in the Python package DEWPython. DEWPython comprises a replication of the DEW spreadsheet in the Python programming language that operates functionally similar, but additionally builds in a dictionary of minerals/aqueous species under *psat* conditions, produces high-resolution datasets without the runtime limitations of Excel, fixes coded errors in DEW, and integrates SUPCRTBL inline. We've also provided the options to optimize/expedite running reactions in DEW. Furthermore, DEWPython features an interactive user interface and extensive documentation that allow the user to determine exactly where the outputs come from instead of feeding through a black box of Visual Basic integrated macros.

Generally, the results from SUPCRT are consistent with the DEW results in the PT regions where they overlap. However, in several cases (e.g., the dissolution of aqueous calcium carbonate and the dissolution of methanol; Figs. 3 and 5) the results for ΔG and log K differ at high pressures and temperatures.

The discrepancies in specific reactions between SUPCRT and DEW could originate from different data sources for specific aqueous organic species in the former. Given the composition of the SUPCRT slop16 (a sequential access database containing thermodynamic data for organic, ionic, and mineral species) from multiple sources (e.g., Canovas and Shock, 2016, which is not contained in DEW), it is plausible that internal inconsistencies exist within the database that cause different end behaviors under *psat* conditions. Similarly, because DEW obtains some species' thermodynamic data from different sources than SUPCRT models, it is highly plausible they will eventually encounter differences for these species.[1] Additionally, there is a slight (< 0.08 MPa) difference in the manner in which pressures are calculated in the *psat* for both DEW and for SUPCRT which results from a difference in equations used between the two programs. Species with a greater pressure dependence could cause a difference in the end result as well.

The Python integration of DEW provides a robust method to determine the possible organic species that occur in the oceans of icy worlds. The favorability of organic species can then be combined with data from the upcoming Europa Clipper or Dragonfly missions to validate and refine the model assumptions for all icy ocean worlds. We find that the different temperature profiles of different ocean worlds shift the stability of the chemical reactions tested here and yield markedly different amounts of minerals and products (§3.2). For example, for a fixed concentration of dissolved calcium and carbonate ions, the production of calcite is about four orders of magnitude higher within Ceres than within Europa at a fixed temperature, and at least three orders of magnitude higher within Titan than within Europa (Fig. 5). In addition, the source of ethane in Titan's atmosphere and lakes is unlikely to be the dissolved methane in Titan's ocean, since ethane appears to be unstable relative to methane under the relevant temperature and pressure conditions (Fig. 4). Finally, methanol is thermodynamically stable relative to its dissociation products in Titan's and Europa's oceans, signifying that it may be present in the oceans, where it could lower the liquidus temperature of water, in line with experimental predictions from Dougherty et al. (2018).

However, there is still room for refinement of DEWPython and extension of its applications. The first main addition to DEWPython will be the addition of more organic species to the aqueous species text file. A first step here is the inclusion of the organic species outlined in Shock, 2016 for the evaluation of reactions present in the citric acid cycle. DEWPython could also be used in conjunction with FREZCHEM (Marion et al., 2010) to further model reactions at lower temperatures.

[1] Full lists of sources for both DEW and SUPCRT are available in the DEW 2019 model and slop16 file respectively.

An extension of these added organic molecules could convolve the use of DEW to evaluate the thermodynamic favorability of reactions relating to different chondrite compositions. A simplified version of DEW is implemented in the Gibbs energy minimization code Perple_X (Connolly, 2017, 2009; Connolly and Galvez, 2018), but it does not include the database extensions we have added here. Adding the thermodynamic database solutes, organics and minerals to such a code would be a useful endeavor.

Further refinement of the DEWPython package will also be helpful. There are specifically several refinements on the documentation/organization side of DEWPython that will make the package more user-friendly, such as the implementation of a graphic user interface, or standardization of variable names/printouts of variable names.

Having the values for the change in free energy provides a robust ground truth that could potentially provide the basis for a neural network. Therefore, another potential application of DEWPython is for the machine learning (ML) analysis of the free energy values of these reactions to the known species compositions.

## Acknowledgements

The Authors would like to acknowledge The Caltech Associates for their contribution to the SURF program which allowed this research to be completed. We thank Jessica Weber (JPL) and H. J. Cleaves (ELSI) for helpful discussions. MMD was partially supported by NASA grants NNH18ZDA001N-HW and NNH19ZDA001N-ECA. A part of the research was carried out at the Jet Propulsion Laboratory, California Institute of Technology, under a contract with the National Aeronautics and Space Administration (80NM0018D0004). 

## Computer Code Availability

Name of Code: DEWPython

Name of Developer: Andrew Chan

Contact Address: 7994 E. Bayaud Ave. Denver, CO. 80230.

Phone: +1 (303) 478-3334

Email: achan1861@gmail.com

Year first available: 2020

Hardware required: computer capable of running Python scripts

Software required: Python: 3.6.10

Packages required: pandas: 0.25.3, numpy: 1.18.5, json: 2.0.9, matplotlib: 3.2.2

Program language: Python

Program Size: 8.82 MB

Access: https://github.com/chandr3w/DEWPython